\documentstyle[12pt,epsfig]{article}
\renewcommand{\thefootnote}{*}
\global\advance\count 0 by -1
\begin{document}
\rightline{July 2002}
\vskip 2.6cm
\centerline{
{\Large \bf Does mirror matter exist?\footnote{
For a postscript version containing all figures
see $http://www.ph.unimelb.edu.au/^{\small \sim}foot/rev.ps$}
}}
\vskip 2.5cm
\centerline{
{\large Robert Foot}
}
\vskip 0.9cm
\centerline{
School of Physics, The University of Melbourne}
\centerline{
Victoria 3010 Australia}
\centerline{E-mail: foot@physics.unimelb.edu.au}
\vskip 2.0cm
\centerline{Abstract}

\vskip 0.6cm

One of the most fascinating ideas coming from particle physics is
the concept of mirror matter. Mirror matter is a new form of
matter which is predicted to exist if mirror symmetry is
respected by nature. At the present time evidence that mirror matter 
actually exists is in abundance, coming from a range of observations
and experiments in astronomy, particle physics, meteoritics and
planetary science.

\newpage

\section{Introduction}

\vskip -0.1cm

Prior
\footnote{The aim of this article is
to provide a fairly non-technical and up to date
review of the motivations for mirror matter and the
evidence for its existence. For a more detailed 
and (hopefully) entertaining exposition see the recent book\cite{book}.}
to 1957 scientists believed that mirror reflection
symmetry was respected by the interactions of the fundamental particles.
Why? Perhaps because the other geometrical symmetries such as
rotations and translations in space and time were observed
to be good symmetries, it seemed natural that mirror reflection symmetry
should be a good symmetry too. Furthermore, no experiment up until
1957 had indicated that these symmetries were anything but
good symmetries of nature. 

In 1956 Lee and Yang\cite{ly6} proposed that the interactions
of the fundamental particles were not mirror reflection
invariant. They suggested that this could explain
some known puzzles and proposed some new experiments 
to directly test the idea. Subsequently Madam C.S.Wu and
collaborators dramatically confirmed that the interactions
of the known particles were not mirror symmetric, just as
Lee and Yang had suspected.

Today, it is widely believed that mirror symmetry is in fact
violated in nature. God -- it is believed -- is left-handed.
Actually, though, things are not so clear.
What the experiments in 1957 and subsequent
experiments have conclusively demonstrated is that
the {\it known} elementary particles behave in a way which
is not mirror symmetric. 
The weak nuclear interaction is the culprit, with the
asymmetry being particularly striking for the weakly
interacting neutrinos.
For example, today we
know that neutrinos only spin with one orientation. If one
was coming towards you it would be spinning like a left-handed
corkscrew. Nobody has ever seen a right-handed neutrino.

The basic geometric point is illustrated in the following diagram:
\vskip 0.5cm
\centerline{\epsfig{file=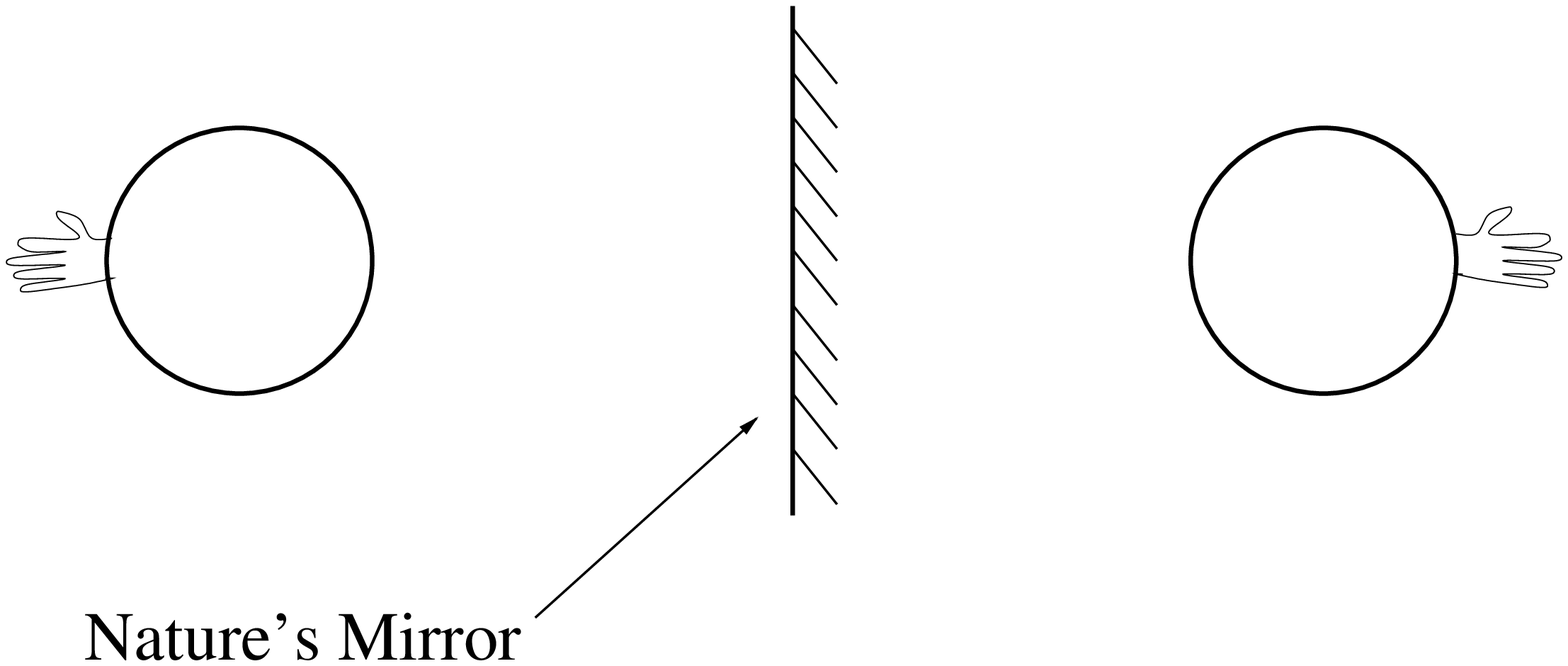,width=8.1cm}}
\vskip 0.35cm
\noindent 
The left-hand side of this figure represents the interactions
of the known elementary particles. The forces are
mirror symmetric like a perfect sphere, except for
the weak interaction, which is represented as
a left hand. Also shown is nature's mirror - the vertical
line down the middle. Clearly, the reflection is not the same as
the original, signifying the fact that the 
interactions of the {\it known} particles are not mirror symmetric.
If there were a right hand as well
as a left hand then mirror symmetry would be
unbroken.
\vskip 0.45cm
\centerline{\epsfig{file=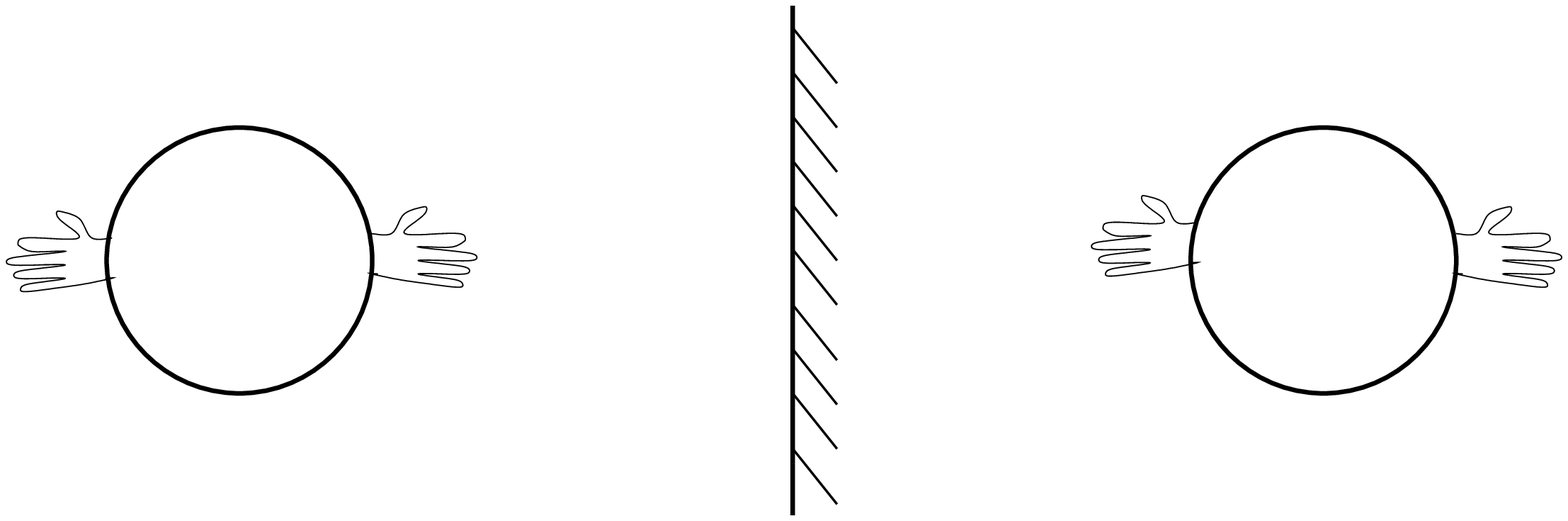,width=7.7cm}}
\vskip 0.5cm
\noindent
However, this doesn't correspond to nature since no right-handed
weak interactions are seen in experiments (this is precisely what the 
experiments in 1957 and subsequently have proven).

There are two remaining
possibilities: We can either chop the hand off -- but this is too
drastic and is therefore not shown. It corresponds to having no weak 
interactions at all, again in disagreement with observations.
This last possibility is the most subtle and consists
of adding an entire new figure with the
hand on the other side. Everything is doubled even the symmetric part,
which is clearly mirror symmetric as indicated in
the following diagram:
\vskip 0.5cm
\centerline{\epsfig{file=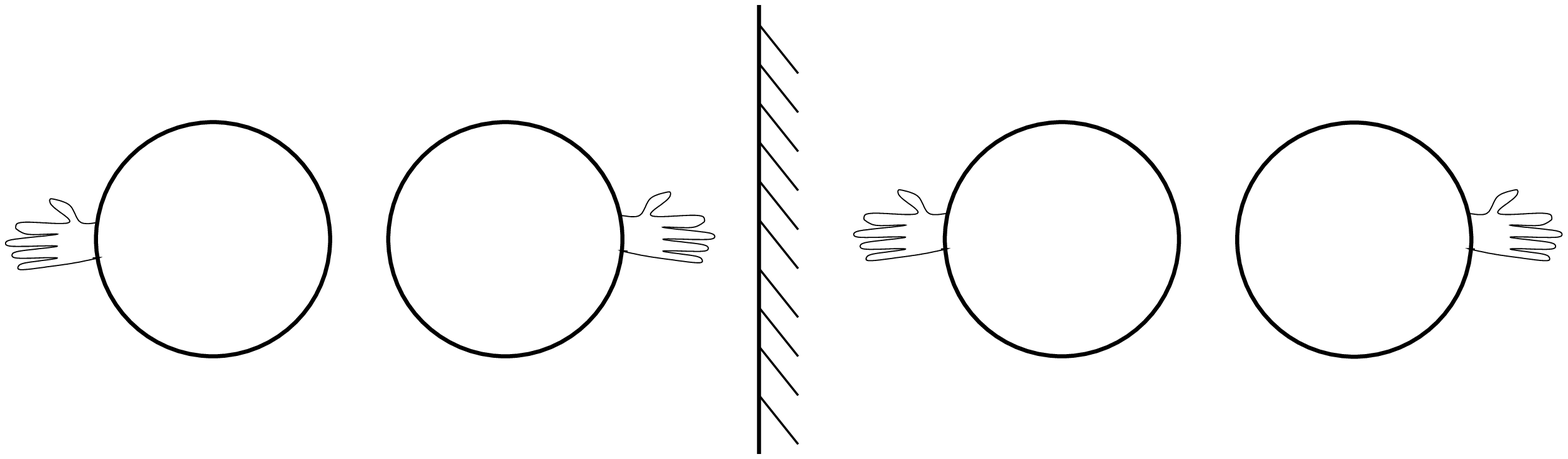,width=9.4cm}}
\vskip 0.5cm
\noindent 
What this figure corresponds to is a complete doubling 
of the number of particles. For each type of particle, such
as electron, proton and photon, there is a mirror twin. 
Where the ordinary particles favor the left hand, the
mirror particles favor the right hand. If such particles
exist in nature, then mirror symmetry would be exactly conserved
(we denote the mirror particles with a prime).
\vskip 0.7cm
\centerline{\epsfig{file=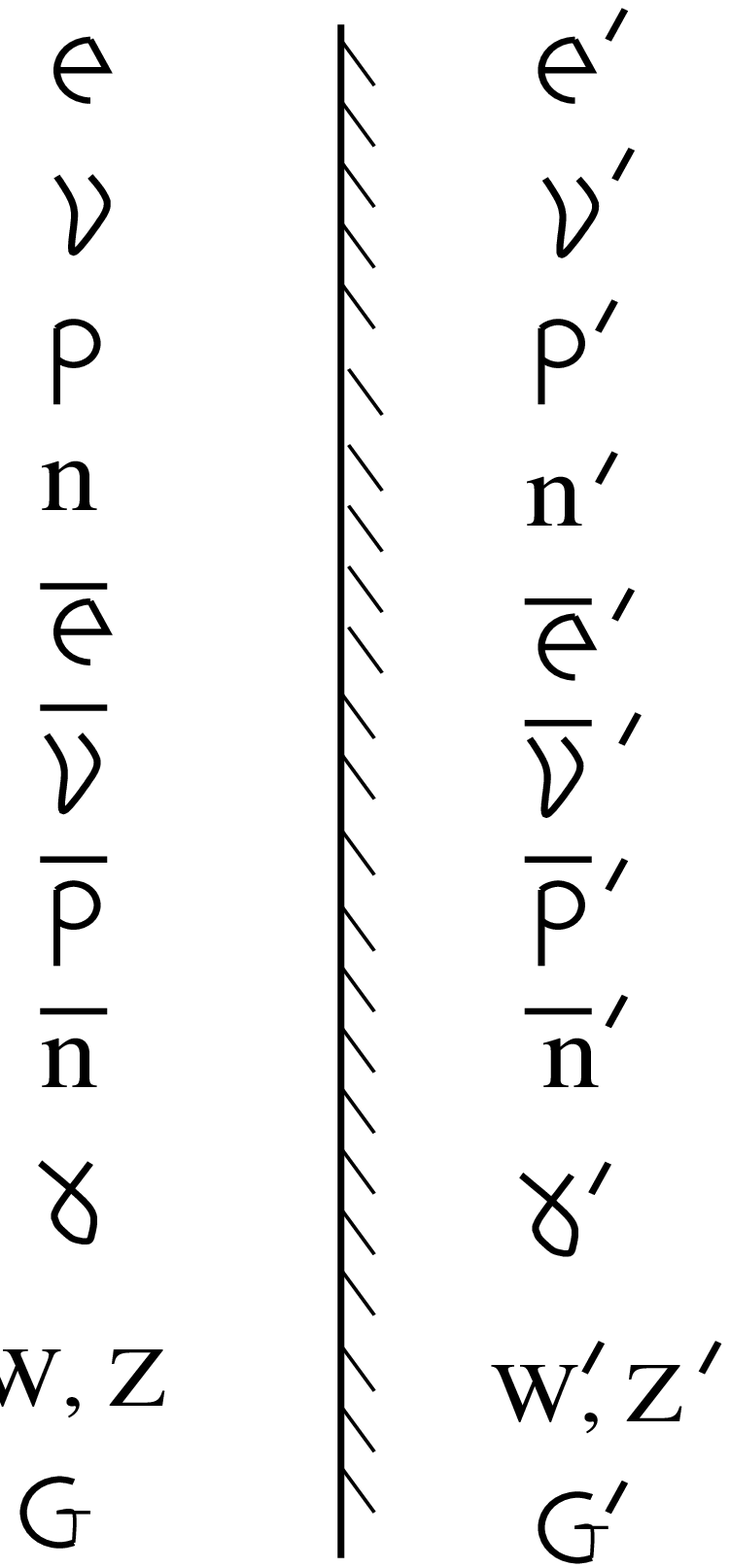,width=3.4cm}}
\vskip 0.5cm
\noindent
As will be discussed, the mirror particles can
exist without violating any known experiment. 
Thus, the correct statement is that the experiments in
1957 and subsequently have only shown that the interactions
of the {\it known} particles are not mirror symmetric, they
have not demonstrated that mirror symmetry is broken in nature.

While many people regard the possible existence  
of mirror particles as highly speculative, it could also be
argued that the assumption of broken mirror symmetry is
equally speculative. Clearly, it is not possible
to figure out which path is chosen by nature on the basis
of pure thought. What really needs to be
done is to understand the experimental implications of the
existence of mirror particles and find out whether such
things could describe our Universe. 

The mirror partners have the same mass as their ordinary
counterparts, which is reminisant of anti-particles.
However, there is a crucial difference. Unlike anti-particles,
the mirror particles interact with ordinary particles predominately 
by gravity only. The three non-gravitational forces act on
ordinary and mirror particles completely separately [and with
opposite handedness: where the ordinary particles are
left-handed, the mirror particles are right-handed].
For example, while ordinary photons 
interact with ordinary matter (which is
just the microscopic picture of the electromagnetic force),
they {\it do not} interact with mirror matter\footnote{
Actually, as will be explained in a moment, it is possible
for small {\it new} forces to exist which connect the ordinary and mirror
particles together. For the purposes of this introductory paragraph,
this possibility is temporarily ignored.
}.
Similarly, the `mirror image' of this statement must also hold,
that is, the mirror photon interacts
with mirror matter but does not interact with ordinary
matter.  The upshot is that we cannot see mirror photons
because we are made of ordinary matter. The mirror 
photons would simply pass right through us
without interacting at all! 

The mirror symmetry does require though that the mirror photons 
interact with mirror electrons and mirror protons in 
exactly the same way in which ordinary photons interact
with ordinary electrons and ordinary protons.
A direct consequence of this is that a mirror atom made
from mirror electrons and a mirror nucleus,
composed of mirror protons and mirror neutrons 
can exist. In fact, mirror matter made
from mirror atoms would also exist with
exactly the same internal properties as ordinary
matter, but would be completely invisible to us! 
Clearly, if there was a negligible amount of mirror matter
in our solar system, we might hardly be aware
of its existence at all. Thus, the {\it apparent}
left-right asymmetry of the laws of nature may
be due to the preponderance of ordinary
matter in our solar system rather than due to a fundamental
asymmetry in the laws themselves.

While this is all very interesting, the most remarkable
thing of all is that there is now a range of evidence
actually supporting the mirror matter theory:
\vskip 0.15cm
\noindent
%1) Mirror matter (e.g. mirror hydrogen composed of mirror protons
%and mirror electrons) should exist in the Universe and would appear
%to us as Dark Matter\cite{kb}.
%Specifically, mirror stars, mirror planets, mirror comets 
%and mirror gas/dust should all exist.
1) It predicts the existence of mirror matter in the Universe.
Mirror matter would be invisible, making its presence felt by its
gravitational effects. Remarkably, there is a large
body of evidence for such invisible `dark' matter. There is
also specific evidence that mirror stars have been observed from
their gravitational effects on the bending of light from background
stars. 

\vskip 0.20cm
\noindent
On the quantum level, small new fundamental 
interactions connecting ordinary and mirror matter are possible.
Various theoretical constraints suggest only a few possible types of
interactions: photon-mirror photon kinetic mixing and
neutrino-mirror neutrino mass mixing\cite{flv,flv2}.
Such non-gravitational forces are extremely
important and open up new ways in which to test the theory:
\vskip 0.15cm
\noindent
2) Orthopositronium should have a shorter effective lifetime 
(in a ``vacuum" experiment) than predicted
due to the effects of photon - mirror photon kinetic mixing\cite{gl,fg}.
\vskip 0.15cm
\noindent
3) If there are small mirror matter bodies in our solar system then
this would lead to a new class of cosmic impact events when
such bodies strike the Earth. Characteristics of such 
impacts will be the lack of ordinary fragments and other anomalous features.
Such impacts will leave mirror matter embedded in the ground which
could potentially be extracted and purified (assuming that the 
small photon-mirror photon kinetic mixing force
exists, which is strong enough to oppose the Earth's feeble gravity).
\vskip 0.15cm
\noindent
4) If there is some remnant mirror hydrogen gas in our solar system, then
spacecraft will experience a drag force and slow down.
\vskip 0.15cm
\noindent
5) If neutrinos have mass then oscillations between ordinary
and mirror neutrinos can occur. Such effects could show up in
neutrino physics experiments.
\vskip 0.1cm

At the present time there is interesting experimental/observational
evidence supporting all five of these predictions.
We now describe this evidence in more detail.

\section{Implications of the mirror world for cosmology}
\vskip 0.15cm
\noindent
There is strong evidence for a large amount 
of dark matter in the Universe. Flat rotation curves of 
stars in spiral
galaxies imply that there must exist invisible halos in
galaxies such as our own Milky Way.
\vskip 0.5cm
\centerline{\epsfig{file=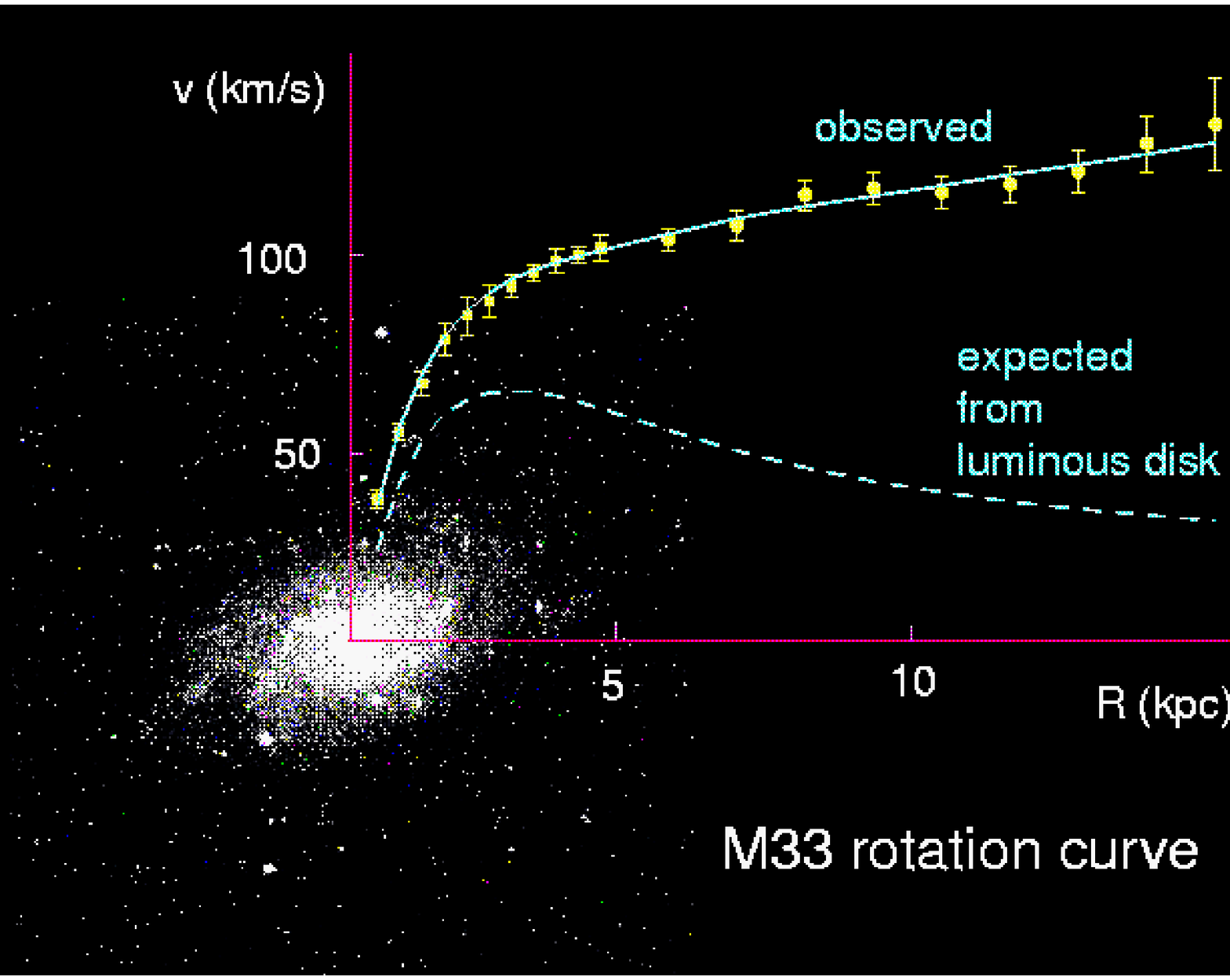,width=9.7cm}}
\vskip 0.7cm
\begin{quotation}
\noindent 
{\small The observed orbiting velocities of stars in the spiral
galaxy M33 superimposed on its optical image. The horizontal
axis is the distance from the center of the galaxy in kiloparsecs.
The poor agreement between the expected velocities and the
actual ones provides strong evidence for invisible `dark matter'.
[Credit for this figure belongs to Ref.\cite{fig}].}
\end{quotation}
\vskip 1.0cm
\noindent
There is also strong
evidence that this dark matter must be something exotic:
ordinary baryons simply cannot account for it\cite{freeze}.
Mirror baryons are necessarily stable and dark (because the coupling 
of mirror matter to ordinary photons is necessarily very small)
%\footnote{ A very small coupling between ordinary and mirror photons is 
%allowed and suggested by experiments measuring the orthopositronium
%lifetime (see next section) but this interaction
%is too small to make mirror matter observable.  \cite{ig}.  })
and are a very natural candidate for the inferred dark matter in the 
Universe\footnote{
Note that various observations indicate
that the amount of dark matter is more than 10 times
the amount of ordinary matter in the Universe. This is not a problem for
the mirror matter theory because a mirror symmetric microscopic
theory does not actually imply 
equal numbers of ordinary and mirror atoms in the Universe.
The point is that the initial conditions need not be
mirror symmetric. The Universe could have been created with
more mirror matter than ordinary matter; alternatively it
is possible that the macroscopic asymmetry was generated during the 
early evolution of
the Universe.}.
This has been argued for some time by Sergei Blinnikov and 
Maxim Khlopov\cite{kb}. 
%(see also the recent reviews in \cite{bd}).
In fact dark matter made of mirror matter would have the property of
clumping into compact bodies such as mirror stars. This leads naturally
to an explanation
%\cite{exp} 
for the mysterious Massive Astrophysical Compact Halo Objects (or MACHO's)
inferred by the MACHO collaboration.
%\cite{macho}.

This collaboration has been studying the nature of halo dark 
matter by using the gravitational microlensing technique.
This Australian-American experiment has collected 5.7 years
of data and provided statistically strong evidence for
dark matter in the form of invisible star sized objects which
is what you would expect if there was a significant
amount of mirror matter in our galaxy.
%\cite{exp}.
The MACHO collaboration \cite{macho} 
have done a maximum likelihood analysis which implies
a MACHO halo fraction of $20\%$ for a typical halo model
with a $95\%$ confidence interval of $8\%$ to $50\%$.
Their most likely MACHO mass is between $0.15 M_{\odot}$ and
$0.9M_{\odot}$ depending on the halo model.
These observations
are consistent with a mirror matter halo because
the entire halo would not be expected to be
in the form of mirror stars. Mirror gas and dust would
also be expected because they are a necessary consequence
of stellar evolution and should therefore significantly
populate the halo. 

If mirror matter exists in our galaxy, then
binary systems consisting of ordinary and mirror matter
should also exist. While systems containing approximately
equal amounts of ordinary and mirror matter are unlikely due
to e.g. differing rates of collapse for
ordinary and mirror matter (due to different 
initial conditions such as chemical composition, temperature
distribution etc),
systems containing predominately ordinary matter 
with a small amount of mirror matter (and vice versa)
should exist. Remarkably, there is interesting evidence for
the existence of such systems coming from extra-solar
planet astronomy.

In the past few years more than 100 ``extrasolar" planets
have been discovered orbiting nearby stars\cite{www}. 
They reveal their presence because their gravity tugs
periodically on their parent stars leading to
observable Doppler shifts. In one case, the planet HD209458b,
has been observed to transit its star
%\cite{trans} 
allowing for an accurate determination of its size and mass.
One of the surprising characteristics of the extrasolar planets
is that there are a class of large ($\sim M_{Jupiter}$) 
close-in planets (with a typical orbital radius of 
$\sim 0.05 \ A. U.$, that is, about 8 times closer than
the orbital radius of Mercury). Ordinary
(gas giant) planets are not expected to form close to stars because the 
high temperatures do not allow them to form.
Theories have been invented where they form far from the
star where the temperature is much lower,
and migrate towards the star. 

%While such theories are possible,
%there are also difficulties, e.g. the recent discovery
%of a close-in pair of resonant planets\cite{marcy} is unexpected since
%migration tends to make the separation between planets diverge
%(as the migration speeds up as the planet becomes closer to the star).

A fascinating alternative possibility presents itself in the mirror
world hypothesis. The close-in planets may be mirror worlds
composed predominately of mirror matter\cite{plan}. 
They do not migrate significantly, but actually formed close to 
the star which is not a problem for mirror worlds because
they are not significantly heated by the radiation
from the star. This hypothesis can explain the
opacity of the transiting planet HD209458b because mirror
worlds would accrete ordinary matter from the solar
wind which accumulates in the gravitational potential
of the mirror world.
%\cite{footr}. 
It turns out that the effective radius of ordinary matter depends 
relatively sensitively on the mass of the planet, so that this mirror 
world hypothesis can be tested when more transiting planets are discovered.
%\cite{footr}.

If this mirror world interpretation of the close-in planets
is correct then it is very natural that the dynamical
mirror image system of a mirror star with an ordinary planet
will also exist. Such a system would appear to ordinary observers
as an ``isolated" ordinary planet. Remarkably,
such ``isolated" planets have recently been identified
in the $\sigma$ Orionis star cluster\cite{iso}.
These planets have estimated mass of $5-15 M_{Jupiter}$
(planets lighter than this mass range would be too faint to
have been detected at present)
and appear to be gas giants which do not seem
to be associated with any visible star.
Given that the $\sigma$ Orionis cluster is estimated to be less than  
5 million years old, the formation of these ``isolated" planets
must have occurred within this time (which
means they can't orbit faint stellar bodies such
as old white dwarfs). Zapatero Osorio et al\cite{iso}
argue that these findings pose a challenge to
conventional theories of planet formation which are
unable to explain the existence of numerous isolated planetary mass
objects. Thus the existence of these planets is surprising 
if they are made of ordinary matter,
however there existence is natural from the mirror world
perspective\cite{iso2}. 
Furthermore, if the isolated planets
are not isolated but orbit mirror stars then
there must exist a periodic Doppler shift detectable
on the spectral lines from these planets. This
represents a simple way of testing this hypothesis\cite{iso2}.

%There is also recent, tantalizing observational evidence
%for mirror matter from another source: A recent
%weak gravitational microlensing study\cite{Erben}
%has apparently discovered an invisible dark concentration of 
%mass in the vicinity of the cluster, Abell 1942.
%A fascinating possibility is that a mirror
%galaxy (or galaxy cluster) containing virtually no 
%ordinary matter has been discovered.
%Further studies should
%help clarify whether this mirror matter interpretation 
%is correct.

\section{Implications of the mirror world for positronium}

\vskip 0.15cm
\noindent
There are only a few possible ways in which ordinary and mirror matter
can interact with each other besides 
gravity, including: photon - mirror
photon kinetic mixing and neutrino - mirror neutrino 
mass mixing (if neutrinos have mass).

Photon - mirror photon kinetic mixing
is described in quantum field theory by the Lagrangian term:
\vspace{-0.1cm}
\begin{equation}
{\cal L}_{int} = \epsilon F^{\mu \nu} F'_{\mu \nu}
\end{equation}
\vspace{-0.1cm}
where $F^{\mu \nu} \equiv \partial^{\mu} A^{\nu} - \partial^{\nu}
A^{\mu}$ is the usual Field strength tensor, and the $F'$ is
the corresponding quantity for mirror photons.
It corresponds to a small coupling between ordinary and
mirror photons. In the language of Feynman diagrams it
is represented by:
\vskip 0.1cm
\centerline{\epsfig{file=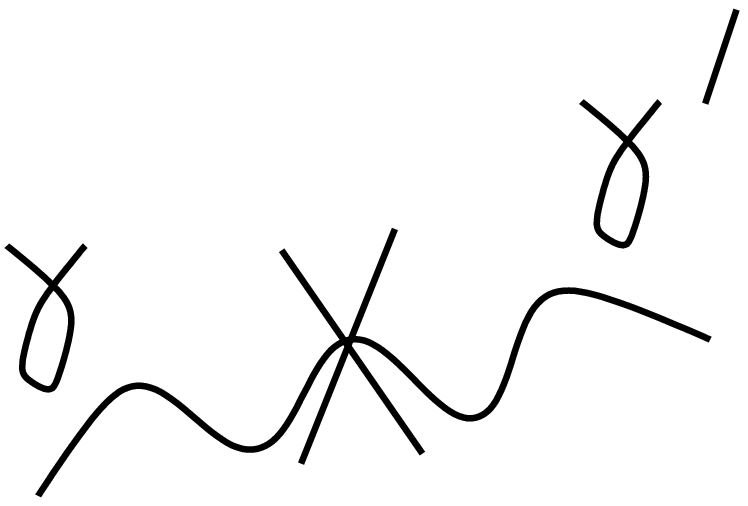,width=4.7cm}}
\vskip 0.8cm
\noindent 
In the absence of any photon-mirror photon kinetic mixing
interaction, an ordinary electron cannot interact with a mirror
electron because ordinary photons do not
interact with mirror electrons (and mirror
photons do not interact with ordinary electrons). However, if
there is a photon-mirror photon kinetic mixing interaction, then 
an ordinary electron can interact with a mirror electron:
\vskip 0.4cm
\centerline{\epsfig{file=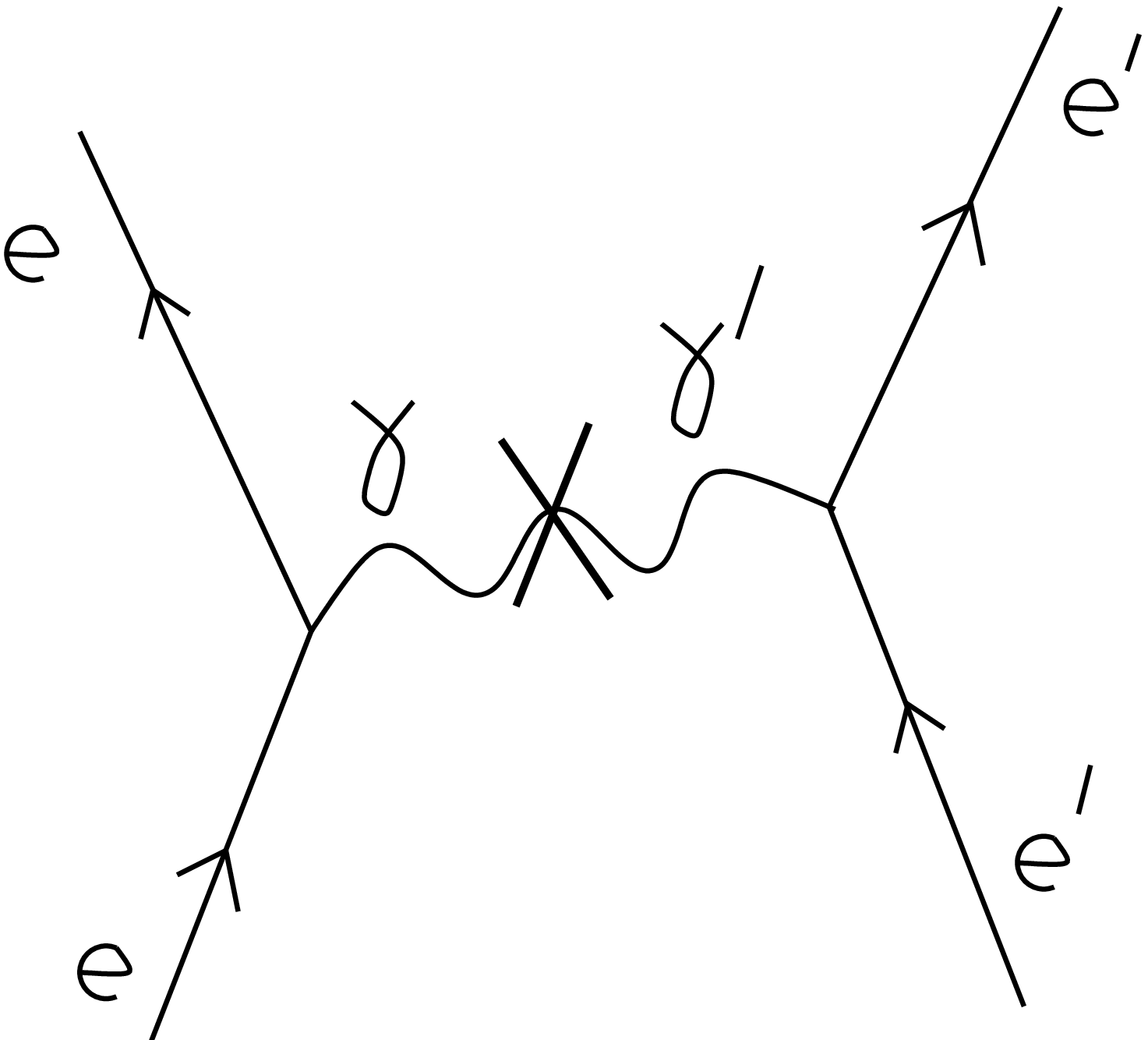,width=6.9cm}}
\vskip 1.2cm
The net effect of the photon-mirror photon mixing is to 
make mirror electrons interact slightly with ordinary electrons
(and similarly for other charged particles).
That is, mirror electrons behave as if they have a tiny 
ordinary electric charge. The size of the effect depends on the 
strength of the photon-mirror 
photon kinetic mixing interaction -- which is characterised by the
parameter $\epsilon$.

It turns out that the most important experimental implication of 
photon - mirror photon kinetic mixing is a rather subtle modification 
of the orthopositronium lifetime\cite{gl}.
Orthopositronium is the bound state
composed of an electron and positron where the spins of
both particles are aligned so that the bound state has spin 1.
The ground state of orthopositronium (o-Ps) decays predominately into
3 photons. The decay rate has been computed in QED leading
to a discrepancy with some of the experimental measurements.
Some of the measurements find a faster decay rate than theoretically
predicted.  This discrepancy has led to a number of experimental
searches for exotic decay modes, including a stringent limit
on invisible decay modes\cite{limit}.

The modification of the lifetime predicted in the mirror
matter theory occurs because the kinetic mixing of the photon
with the mirror photon generates a small off-diagonal
orthopositronium mass leading to oscillations between
orthopositronium and mirror orthopositronium.
The orthopositronium produced in the experiment oscillates
into its mirror partner, whose decays into three mirror photons
are undetected. This effect only occurs in a vacuum experiment
where collisions of the orthopositronium with background
particles can be neglected. Collisions with background
particles will destroy the quantum coherence necessary
for oscillations to occur. Thus, experiments with large
collisions rates remain unaffected by kinetic mixing
and the lifetime of orthopositronium will be the same
as predicted by QED. Experiments in vacuum on the other hand,
should show a slight increase in the decay rate,
as oscillations into mirror orthopositronium and their subsequent
invisible decays effectively reduce the number of orthopositronium
states faster than QED predicts.   

%\vskip 0.8cm
%\centerline{{\Large Orthopositronium.gif Figure here}}
%\vskip 0.8cm

The two most accurate experimental results, normalized to the
theoretical QED 
\linebreak
prediction\cite{afs} are given in the table 
below 
\footnote{A third experiment with gas\cite{aa2} also has
an anomalously high decay rate, however there appears to be
large systematic uncertainties because the orthopositronium
is not thermalized in this experiment\cite{therm,tok}.}
\vskip 0.5cm

{\begin{center}
\begin{tabular}{|l|l|l|l|}
\hline
Reference$\;\;\;\;\;\;\;$
&$\Gamma_{oPs}(exp)/\Gamma_{oPs}(theory)$ $\;\;\;$ 
&Method$\;\;\;\;\;\;\;\;\;\;\;\;$
&$\Gamma_{coll}$$\;\;\;\;\;\;\;\;\;\;\;\;$\\
\hline
Ann Arbor\cite{aa1}&$1.0012\pm 0.0002$&Vacuum Cavity&$\sim (3-10)\Gamma_{oPs}$\\
%Ann Arbor\cite{aa2}&$7.0514\pm 0.0014$&Gas&$\sim 10^3 \Gamma_{oPs}$\\
Tokyo\cite{tok}&$1.0000 \pm 0.0004$&Powder&$\sim 10^4 \Gamma_{oPs}$\\
\hline
\end{tabular}\end{center}}
\vskip 0.2cm
\noindent
\renewcommand{\thefootnote}{**}
Thus, we see that the Tokyo experiment agrees with the QED prediction
while the Ann Arbor vacuum experiment disagrees at about 5 sigma.
These results can be explained in the mirror matter model
by observing that the large collision rate of the
orthopositronium in the Tokyo experiment will
render oscillations of orthopositronium with its mirror
counterpart ineffective\footnote{
The experimental limit\cite{limit} for invisible decay modes
also does not exclude this mirror world oscillation
mechanism because the collision rate of the 
orthopositronium was very high in those experiments.}, 
while the larger decay rate obtained in the vacuum cavity
experiment can be explained
because of the much lower collision rate of orthopositronium
in this experiment allows the
oscillations of ordinary to mirror orthopositronium to take 
effect\cite{fg}. Clearly, an experiment with a larger cavity should
have an even larger effect. Several new experiments are
now being done which will
test for this effect and either confirm (or refute) the
mirror world explanation.
\renewcommand{\thefootnote}{***}

\section{Anomalous Earth impact events} 

\vskip 0.15cm
\noindent
There is not much room for a large amount of mirror matter in our
solar system. For example, the amount of mirror matter within the
Earth has been constrained to be less than $10^{-3}
M_{Earth}$\cite{sashaV}. However, we don't know enough about the
formation of the solar system to be able to exclude the existence
of a large number of  Space Bodies (SB) made of mirror matter if
they are small like comets and asteroids. The total mass of
asteroids in the asteroid belt is estimated to be only about
0.05\% of the mass of the Earth. A similar or even greater number
of mirror bodies, perhaps orbiting in a different plane or even
spherically distributed like the Oort cloud is a fascinating 
possibility\footnote{ Large planetary sized
bodies are also possible if they are in distant
orbits\cite{silnem} or masquerade as ordinary planets or moons
by accreting ordinary matter onto
their surfaces.}. 
[Maybe the Oort cloud itself is composed predominately of
mirror matter bodies, see section 5].

If such small mirror matter bodies exist and happen to collide
with the Earth, what would be the consequences?
If the only force connecting mirror matter
with ordinary matter is gravity, then the consequences
would be minimal. The mirror matter space-body would simply pass
through the Earth and nobody would know about it unless
the body was so heavy as to gravitationally affect the motion
of the Earth. However, if there is a photon-mirror
photon transition force as suggested
by the orthopositronium experiments, then
the mirror nuclei 
can interact with the ordinary nuclei, as illustrated on
the following page.
\vskip 0.7cm
\centerline{\epsfig{file=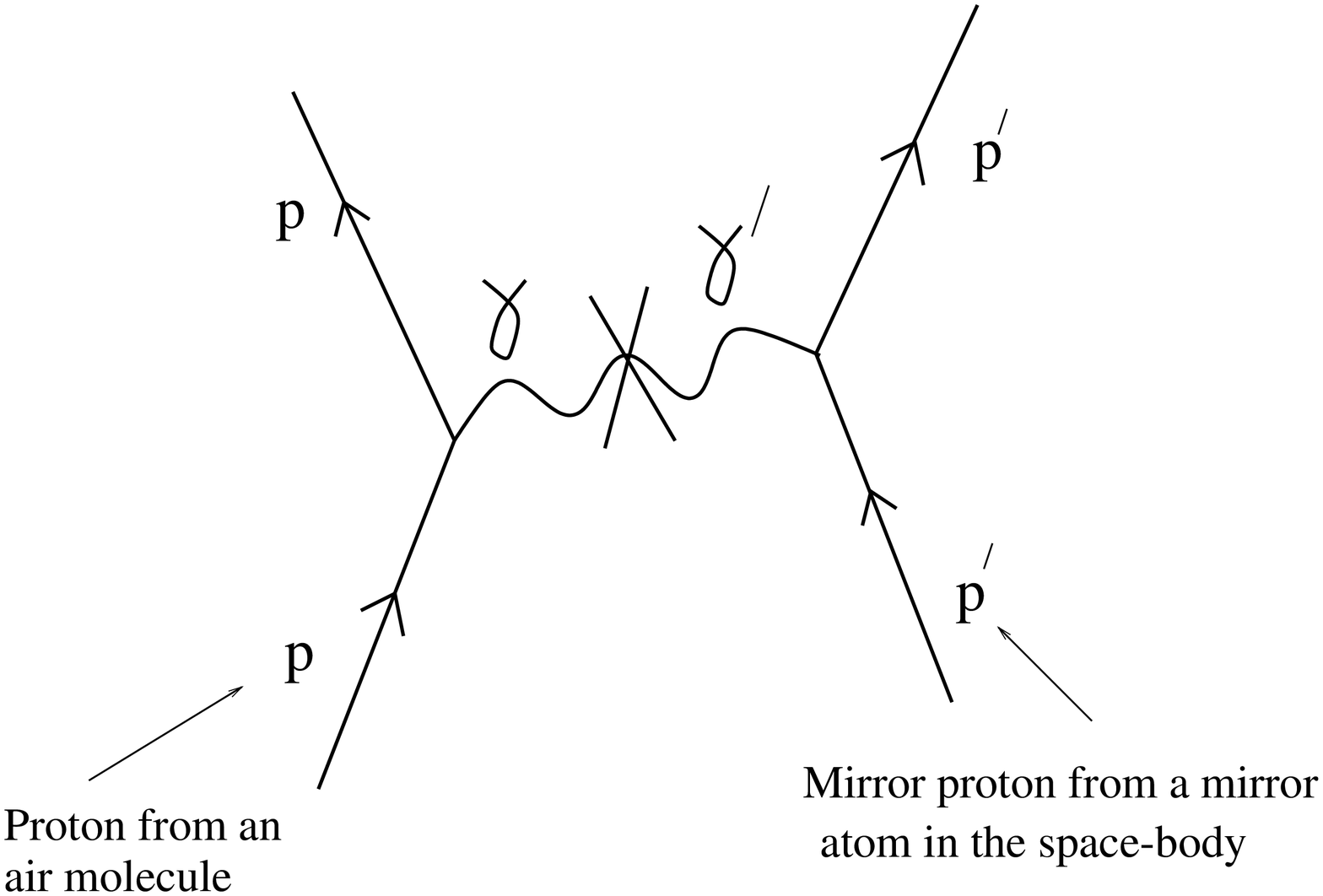,width=9.7cm}}
\vskip 0.5cm
\noindent 
\centerline{{\small Scattering of air as it passes
though the mirror matter space-body.  }}
\vskip 1.0cm
\noindent
In other words, the nuclei of the mirror atoms of the space-body 
will undergo Rutherford scattering with the nuclei of the 
atmospheric nitrogen and oxygen atoms. In addition, ionizing 
interactions can occur (where electrons are removed from the atoms)
which can ionize both the mirror atoms of the space-body and also
the (ordinary) atmospheric atoms.  
This would make the mirror matter space-body effectively visible 
as it plummets to the surface of our planet. 
\renewcommand{\thefootnote}{*}

The rate at which the kinetic energy of a space-body
composed of mirror matter
loses energy through the air depends on a number
of factors, including, the strength of the photon-mirror
photon transition force ($\epsilon$), the chemical
composition of the space-body, its initial velocity
and its size and shape.
We could estimate the initial velocity of the
space-body by observing that the velocity of the Earth
around the Sun is about 30 km/s. The space-body should
have a similar velocity so that depending on its direction,
the relative velocity of the space-body when viewed from
Earth would be expected to be between about 11 and 70 km/s
\footnote{
The minimum velocity of a space-body as
viewed from Earth is not zero because of the
effect of the local gravity of the Earth. 
It turns out that the minimum velocity of a space-body
is about 11 km/s, for a body in an independent orbit
around the sun (and a little less if there happened
to be a body in orbit around the Earth).
}.

It turns out that for 
$\epsilon \stackrel{>}{\sim} 10^{-8}$ (which includes the value
$\epsilon \simeq 10^{-6}$
suggested by the orthopositronium experiments)
the air molecules typically undergo many collisions within
the mirror matter space-body. One important consequence
of this is that the air resistance (or atmospheric `drag force') 
of the mirror matter space-body is actually
roughly the same as if it were made of ordinary matter.
The air resistance will `stop' a space-body in the
atmosphere if it is small in size (roughly less
than 10 meters in diameter) leaving it with a free fall
velocity of $\sim 0.3$ km/s. 
On the other hand, for a large body, much
bigger than 10 meters in diameter, it will not lose
much of its comic velocity in the atmosphere if it remains
intact. 

There are important differences between impacting ordinary
and mirror matter space-bodies.  A mirror body
would heat up {\it internally}
from the interactions of the atmospheric atoms which would
penetrate within the space-body.
This is quite unlike that of an ordinary body which would heat
up only from the friction on the surface.
This means that small mirror matter bodies can have initially
low rates of ablation (i.e. surface melting) and therefore be 
quite dim at high altitudes.
If they are made of non-volatile mirror material (such
as mirror iron) they can
reach the ground without melting and vaporizing in
the atmosphere. Remarkably such types of events are actually
seen in nature.

\vskip 0.3cm
\noindent
{\large \it Some examples of anomalous small fireballs}
\vskip 0.3cm

There are many reported examples of atmospheric
phenomena resembling fireballs, which cannot be
due to the penetration of an ordinary meteoroid into the
atmosphere (for a review of bolides, including discussion
of these anomalous events, see Ref.\cite{bol}).
Below we discuss several examples of this strange
class of phenomena.

\vskip 0.2cm
\noindent
(i) {\it The Spanish event -- January 18, 1994.}

\vskip 0.2cm

On the early morning of 1994 January 18, a very bright luminous
object crossed the sky of Santiago de Compostela, Spain. This
event has been investigated in detail in Ref.\cite{docobo}. The
eye witnesses observed the object to be low in altitude and
velocity ($1$ to $3$ km/s). Yet, an ordinary body penetrating
deep into the atmosphere should have been quite large and luminous
when it first entered the atmosphere at high altitudes with large
cosmic velocity (between $11$ and $70$  km/s). An ordinary body
entering the Earth's atmosphere at these velocities always
undergoes significant ablation as the surface of the body melts
and vapourises, leading to a rapid diminishing of the bodies size
and also high luminosity as the ablated material is heated to high
temperature as it dumps its kinetic energy into the surrounding
atmosphere. Such a large luminous object would have an estimated
brightness which would supersede the brightness of the Sun,
observable at distances of at least $500 \ $km. Sound
phenomena consisting of sonic booms should also have
occurred. Remarkably neither of these two expected
phenomena were observed for this event. The authors of
Ref.\cite{docobo} concluded that the object could {\it not} be a
meteoric fireball.

In addition, within a kilometer of the projected end point of the
``object's'' trajectory a ``crater'' was later
discovered. The ``crater'' had dimensions 
$29$ m $\times 13$ m and $1.5$ m deep. At the crater site, full-grown
pine trees were thrown downhill over a nearby road. Unfortunately,
due to a faulty telephone line on the $17^{th}$ and $18^{th}$ of
January (the fireball was seen on the $18^{th}$) the seismic
sensor at the nearby geophysical observatory of Santiago de Compostela
was inoperative at the crucial time.  After a careful
investigation, the authors of Ref.\cite{docobo} concluded that the
crater was most likely associated with the fireball event, but
could not definitely exclude the possibility of a landslide.
No meteorite fragments or any other unusual material was
discovered at the crater site.

\vskip 0.2cm
\noindent
(ii) {\it The Jordan event -- April 18, 2001.}

\vskip 0.2cm

On Wednesday $18^{th}$ April
2001, more than 100 people attending a funeral
procession saw a low altitude and low velocity fireball.
In fact, the object was observed to break up into two
pieces and each piece was observed to hit the ground.
The two impact sites were later examined by members
of the Jordan Astronomical Society.
The impact sites showed evidence of energy release
(broken tree, half burnt tree, sheared rocks and
burnt ground) 
\linebreak
%\newpage
% (see {\bf Figure 8a,b}). 
%\vskip -21.7cm
%\centerline{\epsfig{file=fig8a.eps,width=12.7cm}}
.
\vskip 3cm
\centerline{{\Large Figure Jordan1.jpg}}
\vskip 3cm
\centerline{{\Large Figure Jordan2.jpg}}
\vskip 1.0cm
\noindent 
%\vskip -18cm
%\centerline{\epsfig{file=fig8b.eps,width=15.0cm}}
%\vskip -0.6cm
\noindent 
\centerline{{\small Pictures of the Jordan impact site\cite{jas}}}
\vskip 1.2cm
%\newpage
\noindent
but no ordinary crater.  [This may have been
due, in part, to the hardness of the ground 
at the impact
sites].  No meteorite fragments were recovered despite
the highly localized nature of the impact sites and low
velocity of impact. For more of the
remarkable pictures and more details, see the Jordan
Astronomical Society's report\cite{jas}.
As with the 1994 Spanish event (i),
the body was apparently not observed by anyone when it 
was at high altitudes where it should have been very bright.
Overall, this
event seems to be broadly similar to the 1994 spanish
event (i). For the same reasons discussed
in (i) (above) it could not be due to an ordinary meteoric fireball.

If these anomalous events are due to the impact of a
mirror matter space-body, then the implications are
extremely important. In particular there should be large
pieces of mirror matter still lodged in the ground
at the impact sites.  The important point
is that the small photon-mirror photon transition force should be large 
enough to oppose the force of gravity\footnote{
Technically, there are two quite distinct cases, depending
on the sign of $\epsilon$. 
(The orthopositronium experiments do not provide any information on
the sign of $\epsilon$). 
Either the photon-mirror photon
mixing induces a small ordinary electric charge for
the mirror electrons ($\epsilon e$) of the same sign as
the ordinary electrons, or the sign is opposite. 
In the first case, the photon-mirror photon mixing force leads
to electrostatic {\it repulsion} between the mirror atoms (of
the mirror matter fragment) and the ordinary atoms in the earth. 
In the alternative case of negative $\epsilon$, 
there is actually electrostatic {\it attraction} between
ordinary and mirror atoms. In both cases though,
the interactions should be strong enough to
stop a piece of mirror matter from falling through the 
Earth.  However an important difference between
the two cases is that in the first case
($\epsilon > 0$)
the repulsion will cause mirror matter fragments
to remain on or near the surface, largely unmixed
with ordinary matter, while in the second case
($\epsilon < 0$)
the mirror matter will penetrate the earth
(a few metres perhaps)
becoming completely mixed in with ordinary matter,
and releasing energy in the process.
Maybe this was the cause of the energy release evident at
the Jordan impact sites (c.f. above figures). }.
Clearly an important issue is to develop ways of 
testing for the presence of mirror matter at these
sites. Several ideas have been proposed: 
\begin{itemize}
\item
{\it Thermal effects of the mirror matter.}
Mirror matter can absorb heat from the surrounding ordinary
matter (through the photon-mirror photon kinetic mixing interaction)
and radiate it away into invisible mirror photons. This
will effectively cool the ordinary matter, an effect which could
be measured.

\item
{\it The presence of mirror matter fragments embedded in
ordinary matter.}
Mirror matter, although essentially chemically inert, still
has mass. In principle it could be extracted and purified,
although the most efficient way of doing this is not yet
known.

\item
{\it Heavy element bound-states.}
Heavy mirror elements can actually form bound-states
with ordinary heavy 
elements (e.g. $Fe-Fe'$, $Fe-Pb'$ etc)\cite{saifoo}. These would appear as
anomalously heavy elements which could show up in
careful mass spectrometer studies.

\end{itemize}

Finally, let me also mention that a large mirror matter
body may have been responsible for the Tunguska event.
For more details of this mirror matter interpretation of the anomalous
small meteoritic events and also application to
the Tunguska event, See Ref.\cite{foottung,book}.
%\newpage
%({\bf Figure 7}).
%\vskip -6.5cm
%\centerline{\epsfig{file=fig9.eps,width=11.2cm}}

\vskip 2cm
\centerline{{\Large Figure 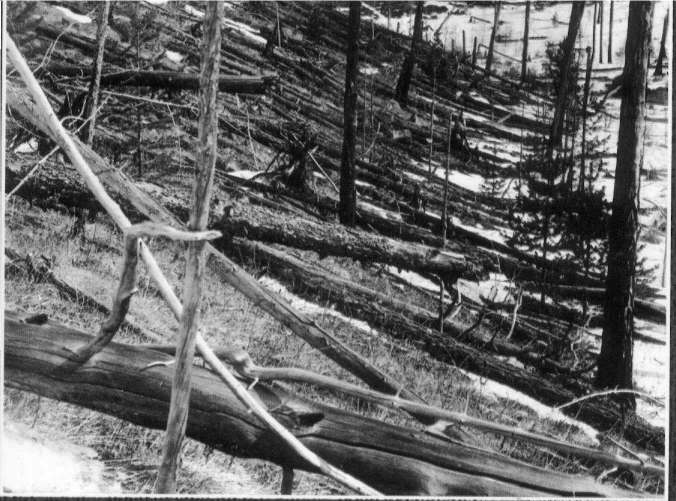}}
\vskip 0.6cm
%\vskip -4.9cm
\centerline{{\small Felled trees as seen by Kulik during his 1928 Tunguska
expedition.} }
\vskip 0.6cm

%\vskip -1.3cm
\noindent 

%%%%%%%%%%%%%%%%%%%%%%%%%
\vskip 1cm
\section{Comets, atmospheric anomalies and pioneer spacecraft}

If mirror matter space-bodes do exist in our 
solar system, then one might expect other
scientific implications. In order not to bore the
reader too much, I will mention only a few of these
things -- and only very briefly.

\subsection{Are comets made of mirror matter?}

Comets are believed to originate from an
approximately spherically symmetric cloud extending out
about half way to the nearest star. This comet cloud, called
the Oort cloud, is reminisant of the dark halo of our galaxy.
Both are largely invisible, are distributed differently to
the `visible' matter, and are also hypothetical. Of course,
this anology is very simplistic and should not be taken very
seriously. Nevertheless, it is also true that
Comets seem to have a number of puzzling features and
are not altogether well understood.
One interesting feature of comets is that 
they seem to contain a very dark nucleus, as shown
in the picture of comet Halley.
\newpage
.
%\vskip -3.2cm
\vskip 3cm
\centerline{{\Large Figure halley.gif}}
\vskip 1cm
%\centerline{\epsfig{file=fig10.eps,width=12.2cm}}
\begin{quotation}
\noindent
{\small  Comet Halley's Nucleus. This picture
was taken by  the spacecraft Giotto. Contrary to prior
expectations, Halley's nucleus is very dark -- one
of the darkest objects in the solar system. Credit:
Halley Multicolour Camera Team, ESA.}
\end{quotation}
\vskip 1.2cm
Its nucleus has an albedo of only 0.03 making it one
of the darkest objects in the solar system -- darker
even than coal! This has led me to wonder whether
the nucleus could be composed predominately of mirror
matter. Of course, pure mirror matter would be
transparent, but if it contained a small admixture
of ordinary matter embedded within, it could appear
opaque and dark. If the ordinary matter had a volatile
component such as water ice, then this would explain
the large head and tail observed when the comet passed
close to the sun. Such a picture would also be
consistent with observations suggesting that many comets
lose a large factor 
(100-1000) in average brightness
after approaching the sun for the first time.
If this interpretation is correct, then comets may
simply become dimmer and dimmer over time eventually
losing all of their volatile ordinary matter component.
They may effectively become invisible. Of course, the
rate that this occurs may depend on many things
such as the proportion of ordinary to mirror matter,
the chemical composition, details of the orbit etc.

Interestingly, a recent study\cite{yyy} has concluded
that many old comets must have either become invisible
or have somehow disintegrated.
The number of cometary remnants (asteroid-like objects) is
100 times less abundant than theoretically expected!\cite{yyy}.
Clearly, this seems to support (or at least, encourage) the 
mirror matter interpretation of the comets. 
Of course, if many comets are predominately made 
of mirror matter then this fits in nicely with the
mirror matter interpretation of the anomalous small fireballs
(and Tunguska event), which was discussed in section 4.
It might also be connected with atmospheric anomalies, as
will now be discussed.

\subsection{Atmospheric anomalies caused by small mirror 
matter space-bodies?}

To explain the anomalous small fireball events 
we require the mirror matter space-body to
survive and hit the ground without completely melting and
vaporizing.  Detailed studies\cite{foottung}
have shown that this is possible for initial velocities near
the minimum ($\sim 11 \ km/s$) and also for non-volatile
mirror matter (such as mirror iron).
A mirror space-body composed of volatile material such
as mirror water ice would
be expected to heat up enough to completely vaporize
in the atmosphere (unless it was very large $\stackrel{>}{\sim} 10$ 
meters in diameter). After vaporizing, the mirror atoms interact
with the air atoms by Rutherford scattering. Although
initially the mirror matter will heat up the ordinary matter
because of its large kinetic energy (since its initial
velocity is at least 11 km/s), after a short time, the mirror matter
will cool the atmosphere. The mirror atoms will draw in heat
from the surrounding ordinary atoms and radiate it away into
mirror photons. Since the mirror atoms are not absorbing mirror
photons from the environment, heat will be lost from the system.
The net effect is a localized rapid cooling 
of the atmosphere which might lead to the formation of unusual clouds
and other strange atmospheric phenomena. Perhaps this might
explain the remarkable observations of
falling ice blocks\cite{ib} and maybe even the observations
of atmospheric `holes'\cite{hole}.

\subsection{Anomalous acceleration of the Pioneer spacecraft explained?}

\vskip 0.15cm
\noindent

Another indication for mirror matter in our
solar system comes from the Pioneer 10 and 11 spacecraft anomalies.
These spacecraft, which are identical in design, were launched in 
the early 1970's with Pioneer 10 going to Jupiter and Pioneer
11 going to Saturn.  After these planetary rendezvous, 
the two spacecraft followed
orbits to opposite ends of the solar system 
with roughly the same speed, which is now about 12 km/s.  
The trajectories of these spacecraft were carefully monitored
by a team of scientists from the Jet Propulsion
Laboratory and other institutions\cite{study}.
The dominant force on the spacecraft is, of course, the gravitational 
force, but there is also another much smaller force
coming from the solar radiation pressure -- that is, 
a force arising from the light striking the surface of the spacecraft.
However, the radiation pressure 
decreases quickly with distance from the sun, and for distances greater
than 20 AU it is low enough to allow for a 
sensitive test for anomalous forces in the solar system. The
Pioneer 11 radio system failed in 1990 when it was about 30
AU away from the Sun, while Pioneer 10 is in better shape
and is about 70 AU away from the Sun (and still transmitting!).

The Pioneer 10/11 spacecrafts are very sensitive probes
of mirror gas and dust in our solar system if 
the photon-mirror photon transition force exists as suggested
by the orthopositronium experiments.
Collisions of the spacecraft with mirror particles
will lead to a drag force which will slow the spacecraft down.
This situation of an ordinary matter body (the spacecraft)
propagating though a gas of mirror particles is
a sort of `mirror image' of a mirror matter space-body
propagating through the atmosphere which was considered in
the previous section.

Interestingly, careful and detailed studies\cite{study} of the motion
of Pioneer 10 and 11 have revealed
that the accelerations of {\it both} spacecrafts are anomalous 
and directed roughly towards the Sun, with magnitude,
$a_p = (8.7 \pm 1.3)\times 10^{-8}\ {\rm cm/s^2}\ $.
In other words, the spacecrafts are inexplicably slowing down!
Many explanations have been
proposed, but all have been found wanting so far.
For example, ordinary gas and dust cannot explain it
because there are rather stringent constraints on
the density of ordinary matter in our solar system
coming from its interactions with the sun's light.
However, the constraints on
mirror matter in our solar system are much weaker
because of its invisibility as far as its interactions with
ordinary light is concerned.

If this anomalous acceleration of the spacecraft
is due to remnant mirror matter gas or dust in our solar
system, then calculations of Ray Volkas and myself\cite{fvpioneer}
suggest a density of mirror matter in 
our solar system of about $\approx 4 \times 10^{-19}\ {\rm g/cm^3}$.
It corresponds to about 200,000 mirror hydrogen atoms 
(or equivalent) per cubic centimetre. If the mirror gas/dust
is spherically distributed with a radius of order 100 AU, then the total
mass of mirror matter would be about that of a small planet ($\approx
10^{-6} M_{sun}$) with only about $10^{-8} M_{sun}$
within the orbit of Uranus, which is about two orders
of magnitude within present limits.  If the configuration
is disk-like rather than spherical, then the total mass of mirror
matter would obviously be even less. The requirement
that the mirror gas/dust be denser than its ordinary counterpart at 
these distances could be due
to the ordinary material having been expelled by solar pressure...

\section{Implications of the mirror world for neutrino physics}
\vskip 0.15cm
\noindent
If neutrinos have mass, then it is possible for ordinary and
mirror neutrinos to oscillate into each other.
Neutrino oscillations are a well known quantum mechanical effect
which arise
when the flavour eigenstates are linear combinations
of 2 or more mass eigenstates. For example, if the
electron and muon neutrinos have mass which mixes
the flavour eigenstates, then in general the
weak eigenstates are orthogonal combinations of mass
eigenstates, i.e.
\begin{equation}
\nu_e = \sin\theta \nu_1 + \cos\theta \nu_2,\
\nu_\mu = \cos\theta \nu_1 - \sin\theta \nu_2.
\label{1}
\end{equation}
A standard result 
%(see e.g. Perkins\cite{per}) 
is that the oscillation
probability for a neutrino of energy $E$ is then
\begin{equation}
P(\nu_\mu \to \nu_e) = \sin^2 2\theta \sin^2 L/L_{osc}, 
\label{osc}
\end{equation}
where $L$ is the distance from the source and $L_{osc} \equiv 4E/\delta
m^2$ is the oscillation length (and natural units
have been used, i.e. $c=h/2\pi=1$).
%\footnote{
%In Eq.(\ref{osc}), $\delta m^2 \equiv m_1^2 - m_2^2$ 
%is the difference in squared
%masses of the neutrino mass eigenstates.}.
If $\sin^2 2\theta = 1$ then the oscillations have the greatest
effect and this is called maximal oscillations.

Anyway, in 1992 Henry Lew, Ray Volkas and I\cite{flv2} found the 
remarkable result that
the oscillations between ordinary and mirror
neutrinos are necessarily maximal which is a direct consequence
of the mirror symmetry.
One way to see this is to note that if neutrinos mix then the
mass eigenstates are non-degenerate and 
necessarily parity eigenstates if
parity is unbroken. Considering the first generation electron neutrino,
$\nu_e$ and its mirror partner, $\nu'_e$, 
the parity eigenstates are
simply $\nu^{\pm} = (\nu_e \pm \nu'_e)/\sqrt{2}$ (since parity
interchanges the ordinary and the mirror particles) and hence
\begin{equation}
\nu_e = {\nu^+ + \nu^- \over \sqrt{2}},\
\nu'_e = {\nu^+ - \nu^- \over \sqrt{2}}.
\end{equation}
Comparing this with Eq.(\ref{1}) we see that
$\theta = \pi/4$ i.e. $\sin2\theta = 1$ and hence maximal mixing!
Thus, if neutrinos and mirror neutrinos have mass and mix together then
the oscillations between the ordinary and mirror neutrinos are
necessarily maximal. 

This result is extremely interesting in view of the accumulated
evidence for neutrino oscillations (for a recent
summary, see Ref.\cite{recent} and references there-in). Three types of
experiments suggest the existence of neutrino oscillations
and actually require the existence of at least one new 
`sterile neutrino' (i.e.  non-interacting neutrino such as a 
mirror neutrino) maximally mixed with one of the known neutrinos.
However, more experimental studies need to be done and are
being done, so the situation in neutrino physics will become
clearer. Of course, it is possible
that there is no detectable mirror world effect for neutrinos,
even if mirror matter exists. While
the mixing is maximal, the oscillation lengths depend on 
the details of the neutrino mass physics. In particular,
in the limit that the mass mixing between ordinary
and mirror neutrinos goes to zero, the effect of the
mirror world must also vanish.

\section{Conclusion}

It is a known fact that almost every plausible symmetry
(such as rotational invariance, translational invariance etc) are
found to be exact symmetries of the particle interactions.
Thus, it would be very strange if the fundamental interactions
were not mirror symmetric. It is a very interesting observation that
mirror symmetry requires the existence of a new form of
matter called `mirror matter', otherwise there is nothing to
balance the left-handed nature of the weak force. Even
more interesting, is the remarkable evidence that mirror matter
actually exists. But, does mirror matter really exist?
I'm not sure, but I would very much like to find out. Maybe the
answer lies in Jordan or maybe it is blowing in the wind\cite{bobd}...

\vskip 0.4cm
\noindent
{\large \bf Acknowledgements}
\vskip 0.1cm
\noindent
It is a pleasure to thank my collaborators,
Sergei Gninenko, Sasha Ignatiev,
Henry Lew, Saibal Mitra, Zurab Silagadze, Ray Volkas and T. L. Yoon.
I would also like to acknowledge very useful correspondence
from Sergei Blinnikov, Zdenek Ceplecha, 
Carl Feynman, Luigi Foschini, John Learned, 
Jesus Martinez-Friaz and Andrei Ol'khovatov.

\vskip 0.6cm
\noindent
{\large \bf References}
%\vskip 0.2cm


\vspace{-1.5cm}

\begin{thebibliography}{99}

\bibitem{book}
{\it Shadowlands, quest for mirror matter in the Universe},
R. Foot, Universal Publishers (2002). For other information
and useful links, see also the mirror matter webpage:
http://www.ph.unimelb.edu.au/~foot/ (or the mirror matter
mirror site, http://www.geocities.com/mirrorplanets).

\bibitem{ly6}
T. D. Lee and C. N. Yang, Phys. Rev. 104, 256 (1956). 

\bibitem{flv}
R. Foot, H. Lew and R. R. Volkas, Phys. Lett. B272, 67 (1991). 
The idea was earlier known to Lee and Yang, 
see the last two paragraphs of Ref.\cite{ly6} 
and a very useful contribution
was also made in I . Kobzarev, L. Okun and I. Pomeranchuk, 
Sov. J. Nucl. Phys. 3, 837 (1966).

\vspace{-1.0mm}
\bibitem{flv2}
R. Foot, H. Lew and R. R. Volkas, Mod. Phys. Lett. A7, 2567 (1992).

\vspace{-1.0mm}
\bibitem{gl}
S. L. Glashow, Phys. Lett. B167, 35 (1986).

\vspace{-1.0mm}
\bibitem{fg}
R. Foot and S. N. Gninenko, Phys. Lett. B480, 171 (2000)
[hep-ph/0003278].

\vspace{-1.0mm}
\bibitem{fig}
D. Roy, physics/0007025.

\vspace{-1.0mm}
\bibitem{freeze}
See e.g. K. Freese, B. Fields and D. Graff, astro-ph/9904401.

\vspace{-1.0mm}
\bibitem{kb}
S. I. Blinnikov and M. Yu. Khlopov, Sov. J. Nucl. Phys.
36, 472 (1982); Sov. Astron. 27, 371 (1983);
E. W. Kolb, D. Seckel and M. S. Turner, Nature 514, 415 (1985);
M. Yu. Khlopov et al., Soviet Astronomy 35, 21 (1991);
H. M. Hodges, Phys. Rev. D47, 456 (1993);
N. F. Bell and R. R. Volkas, Phys. Rev. D59, 107301 (1999)
[astro-ph/9812301];
Z. K. Silagadze, Phys. At. Nucl. 60, 272 (1997) [hep-ph/953481];
S. Blinnikov, astro-ph/9801015; astro-ph/9902305;
R. Foot, Phys. Lett. B452, 83 (1999) [astro-ph/9902065];
Z. Berezhiani, D. Comelli and F. Villante, Phys. Lett. B503, 362 (2001)
[hep-ph/0008105].

\vspace{-1.0mm}
\bibitem{macho}
C. Alcock et al (Macho Collab), Ap.J.542, 281 (2000) [astro-ph/0001272].

\vspace{-1.0mm}
\bibitem{www}
For a review and references on extrasolar planets, see
the extrasolar planet encyclopaedia:
http://cfa-www.harvard.edu/planets/encycl.html

\vspace{-1.0mm}
\bibitem{plan}
R. Foot, Phys. Lett. B471, 191 (1999) [astro-ph/9908276]; B505, 1 (2001)
[astro-ph/0101055].
%R. Foot, Phys. Lett. B505, 1 (2001).

\vspace{-1.0mm}
\bibitem{iso}
M. R. Zapatero Osorio et al., Science 290, 103 (2000).
See also M. Tamura et al., Science 282, 1095 (1998); P. W.
Lucas and P. F. Roche, Mon. Not. R. Astron. Soc. 314, 858 (2000).

\vspace{-1.0mm}
\bibitem{iso2}
R. Foot, A. Yu. Ignatiev and R. R. Volkas, Astropart. Phys.
17, 195 (2002) [astro-ph/0010502].

%\vspace{-1.0mm}
%\bibitem{f} R. Foot, Mod. Phys. Lett. A9, 169 (1994);
%R. Foot and R. R. Volkas, Phys. Rev. D52, 6595 (1995).

%\vspace{-1.0mm} \bibitem{ig}
%R. Foot, A. Yu. Ignatiev and R. R. Volkas, Phys. Lett. B503, 355 (2002).

%\vspace{-1.0mm} \bibitem{bd}
%S. Blinnikov, astro-ph/9911138; A. Dolgov, astro-ph/9910532.

%Z. K. Silagadze, Phys. At. Nucl. 60, 272 (1997);
%S. Blinnikov, astro-ph/9801015;
%R. Foot, Phys. Lett. B452, 83 (1999).

%\vspace{-1.0mm} \bibitem{trans}
%D. Charbonneau et al, Ap J. 529, L15 (2000); G. W. Henry et al, ibid, 
%L41 (2000).

%\vspace{-1.0mm} \bibitem{marcy}
%G. W. Marcy et al, Ap. J (to appear).  \vspace{-1.0mm}

%\vspace{-1.0mm} \bibitem{Erben}
%T. Erben et al, Astronomy and Astrophysics, 355, 23 (2000);
%M. E. Gray et al, astro-ph/0101431.

\vspace{-1.0mm}
\bibitem{limit}
G. S. Atojan, S. N. Gninenko, V. I.
Razin and Yu. V. Ryabov, Phys. Lett. B220,
317 (1989); 
T. Mitsui et al., Phys.  Rev. Lett. 70, 2265 (1993).

\vspace{-1.0mm}
\bibitem{afs}
G. S. Adkins, R. N. Fell and J. Sapirstein, Phys. Rev. Lett. 
84, 5086 (2000).

\vspace{-1.0mm}
\bibitem{aa2}
C. I. Westbrook et al., Phys. Rev. A40, 5489 (1989).

\vspace{-1.0mm}
\bibitem{therm}
M. Skalsey et al., Phys. Rev. Lett. 80,
3727 (1998).

\vspace{-1.0mm}
\bibitem{tok}
S. Asai, S. Orito and N. Shinohara, Phys. Lett. B357, 475 (1995).

\vspace{-1.0mm}
\bibitem{aa1}
J. S. Nico et al, Phys. Rev.  Lett. 65, 1344 (1990).

%\vspace{-1.0mm}
%\bibitem{cern}
%http://pcicarus7.ethz.ch/Positron/
%New Scientist, Vol. 166 (17th June 2000), p36-39.

\vspace{-1.0mm}
\bibitem{sashaV}
A. Yu. Ignatiev and R. R. Volkas, Phys. Rev. D62, 023508 (2000)
[hep-ph/0005125].

\vspace{-1.0mm}
\bibitem{silnem}
Z. K. Silagadze, Acta. Phys. Pol. B32, 99 (2001) [hep-ph/0002255];
R. Foot and Z. K. Silagadze, Acta. Phys. Pol. B32, 2271 (2001)
[astro-ph/0104251]; Z. K. Silagadze, astro-ph/0110161.

\vspace{-1.0mm}
\bibitem{bol}
Z. Ceplecha et al.,
Meteoroids 1998, Astron. Inst., Slovak. Acad. Sci.,
Bratislava, 1999, pp 37-54.

\vspace{-1.0mm}
\bibitem{docobo}
J. A. Docobo, R. E. Spalding, Z. Ceplecha, F. Diaz-Fierros, V. Tamazian
and Y. Onda, Meteoritics \& Planetary Science 33, 57 (1998).

\vspace{-1.0mm}
\bibitem{jas}
http://www.jas.org.jo/mett.html

\vspace{-1.0mm}
\bibitem{saifoo}
R. Foot and S. Mitra, hep-ph/0204256.

\vspace{-1.0mm}
\bibitem{foottung} 
R. Foot, Acta Phys. Pol. B32, 3133 (2001) [hep-ph/0107132];
R. Foot and T. L. Yoon, Acta Phys. Pol. B33, 1979 (2002)
[astro-ph/0203152].

\bibitem{yyy}
H. Levison et al, Science 296 (2002).

\bibitem{ib}
%http://tierra.rediris.es/bloquesdehielo
http://tierra.rediris.es/megacryometeors/index2.html
and J. Martinez-Frias et al, 
Compositional Heterogeneity of
Hailstones: Atmospheric Conditions and Possible Environmental
Implications, Ambio Vol. 30 (2001).
See also 

\noindent
http://www.geocities.com/olkhov/gr1997.htm 
where an interesting catalogue of many strange atmospheric events (of
unknown origin) is given.

\bibitem{hole}
http://smallcomets.physics.uiowa.edu/

\noindent
Note that the idea that mirror matter might be responsible
for these atmospheric holes was suggested to me
by Stephen Heyer.

\vspace{-1.0mm}
\bibitem{study}
J. D. Anderson et al., Phys. Rev. D65, 082004 (2002) [gr-qc/0104064].

\vspace{-1.0mm} 
\bibitem{fvpioneer}
R. Foot and R. R. Volkas, Phys. Lett. B517, 13 (2001) [hep-ph/0108051].

\vspace{-1.0mm}
\bibitem{recent}
R. Foot and R. R. Volkas, hep-ph/0204265.

\bibitem{bobd}
B. Dylan, Public communication.

\end{thebibliography}
\end{document}